\documentclass[aps,prl,reprint]{revtex4-1}
\usepackage{graphicx}
\usepackage{amssymb}
\usepackage{mathtools}
\usepackage{amsmath}
\usepackage{bm}
\usepackage{epstopdf}
\usepackage{xcolor}

\begin{document}

\title{Topological Creutz Ladder in a Resonantly Shaken 1D Optical Lattice}

\author{Jin Hyoun Kang}
\author{Jeong Ho Han}
\author{Y. Shin}\email{yishin@snu.ac.kr}
\affiliation{Department of Physics and Astronomy, and Institute of Applied Physics, Seoul National University, Seoul 08826, Korea \\ and Center for Correlated Electron Systems, Institute for Basic Science, Seoul 08826, Korea}


\begin{abstract}
We report the experimental realization of a topological Creutz ladder for ultracold fermionic atoms in a resonantly driven 1D optical lattice. The two-leg ladder consists of the two lowest orbital states of the optical lattice and the cross inter-leg links are generated via two-photon resonant coupling between the orbitals by periodic lattice shaking. The characteristic pseudo-spin winding in the topologically non-trivial bands of the ladder system is demonstrated using momentum-resolved Ramsey-type interferometric measurements. We discuss a two-tone driving method to extend the inter-leg link control and propose a topological charge pumping scheme for the Creutz ladder system. 
\end{abstract}

\maketitle

Topological phases such as quantum Hall states and topological insulators represent intriguing physics beyond the conventional Landau paradigm of phase transition~\cite{Hasan10,Qi11}. Motivated further by their novel transport properties, the study of topological phases constitutes one of the frontiers in modern condensed matter physics. Ultracold atoms in optical lattices, featuring tunneling amplitude engineering and tunable interaction strength, provide a unique platform for realizing and exploring such exotic topological states~\cite{Goldman16}. Along with the steady development of experimental techniques, many topological model systems have been recently realized, including the Harper--Hofstadter Hamiltonian in 2D rectangular lattices~\cite{Aidelsburger13, Miyake13}, the Haldane model in a 2D hexagonal lattice~\cite{Jotzu14}, and various Hall and topological ladder systems based on additional synthetic dimensions such as internal atomic states~\cite{Mancini15,Stuhl15,Livi16,Kolkowitz17, Song18, Han18} and lattice orbital states~\cite{Kang18}.

Periodic lattice shaking is one of the successful tools for exploring exotic phases in optical lattices. Under periodic temporal modulations of the lattice potential, the system parameters such as tunneling magnitude~\cite{Lignier07,Struck11} and phase~\cite{Struck12} can be coherently manipulated, giving rise to a hopping configuration that is difficult to realize with static schemes. An outstanding example is the Haldane model realized by circularly shaking a 2D hexagonal optical lattice potential to achieve complex next-nearest-neighbor hopping~\cite{Jotzu14,Oka09}. From the perspective of Floquet band engineering, the lattice shaking method has been extensively discussed even in the resonant regime where the driving frequency is high enough to match the energy gap between two bands~\cite{Goldman15}. Such strong orbital hybridization may enable access to a broader range of effective Hamiltonians~\cite{Parker13, Ha15}. In particular, it was anticipated that multi-photon inter-orbital resonant coupling could yield a special route to engineer topological states~\cite{Zheng14,Zhang14}. Thus, it is highly desirable to examine the multifarious scope of Floquet band engineering for the study of topological phases.

In this Letter, we experimentally investigate the effects of two-photon inter-orbital resonant coupling in a periodically driven 1D optical lattice, and demonstrate the realization of a topological Creutz ladder for ultracold fermionic atoms in the shaken lattice system. The Creutz ladder is a cross-linked two-leg ladder system, which has been discussed as a minimal model for 1D topological insulators~\cite{Creutz99, Junemann17}.  In our experiment, the two-leg ladder is formed by the two lowest orbital states, and the cross inter-leg links are generated via the two-photon resonant coupling between orbitals by lattice shaking. Using momentum-resolved Ramsey-type interferometric measurements, we demonstrate the characteristic pseudo-spin winding in the topologically non-trivial bands of the Creutz ladder. We also discuss the extension of the inter-leg link control with two-frequency driving, and propose an experimental scheme for topological charge pumping in the Creutz ladder system.

Our experiment starts by preparing a spin-balanced degenerate Fermi gas of $^{173}$Yb atoms in the $F=5/2$ hyperfine ground state, as described in Ref.~\cite{Lee17}. The total atom number is $\approx 1.5\times 10^5$, and the temperature is $\approx 0.35T_F$, where $T_F$ is the Fermi temperature of the trapped sample. The atoms are adiabatically loaded in a 1D optical lattice, which was formed along the $x$-direction by interfering two laser beams with a wavelength of $\lambda_L = 532$ nm. The lattice spacing and depth are $a=\sqrt{3}\lambda_L/2$ and $V_L=8 E_r$, respectively, where $E_r = h^2/8ma^2 = h\times 3.1$ kHz. The trapping frequencies of the overall harmonic potential are estimated to be $(\omega_x,\omega_y,\omega_z) \approx 2\pi \times (41,61,130)~\text{Hz}$. After the atom loading, we periodically drive the lattice potential by sinusoidally modulating the frequency difference between the two lattice laser beams as $\delta \omega_L(t) = A \sin(\omega t+\varphi)$ [Fig. 1a], which results in lattice site vibrations with amplitude $d=\frac{A}{2\pi \omega}a$. In this work, we investigate the resonant driving cases with $\hbar\omega\approx\epsilon_{sp}$ or $2\hbar\omega\approx\epsilon_{sp}$, where $\epsilon_{sp}$ is the on-site energy difference between the $s$ and $p$ orbitals of the lattice system.

\begin{figure}[t]
\includegraphics[width=8.4cm]{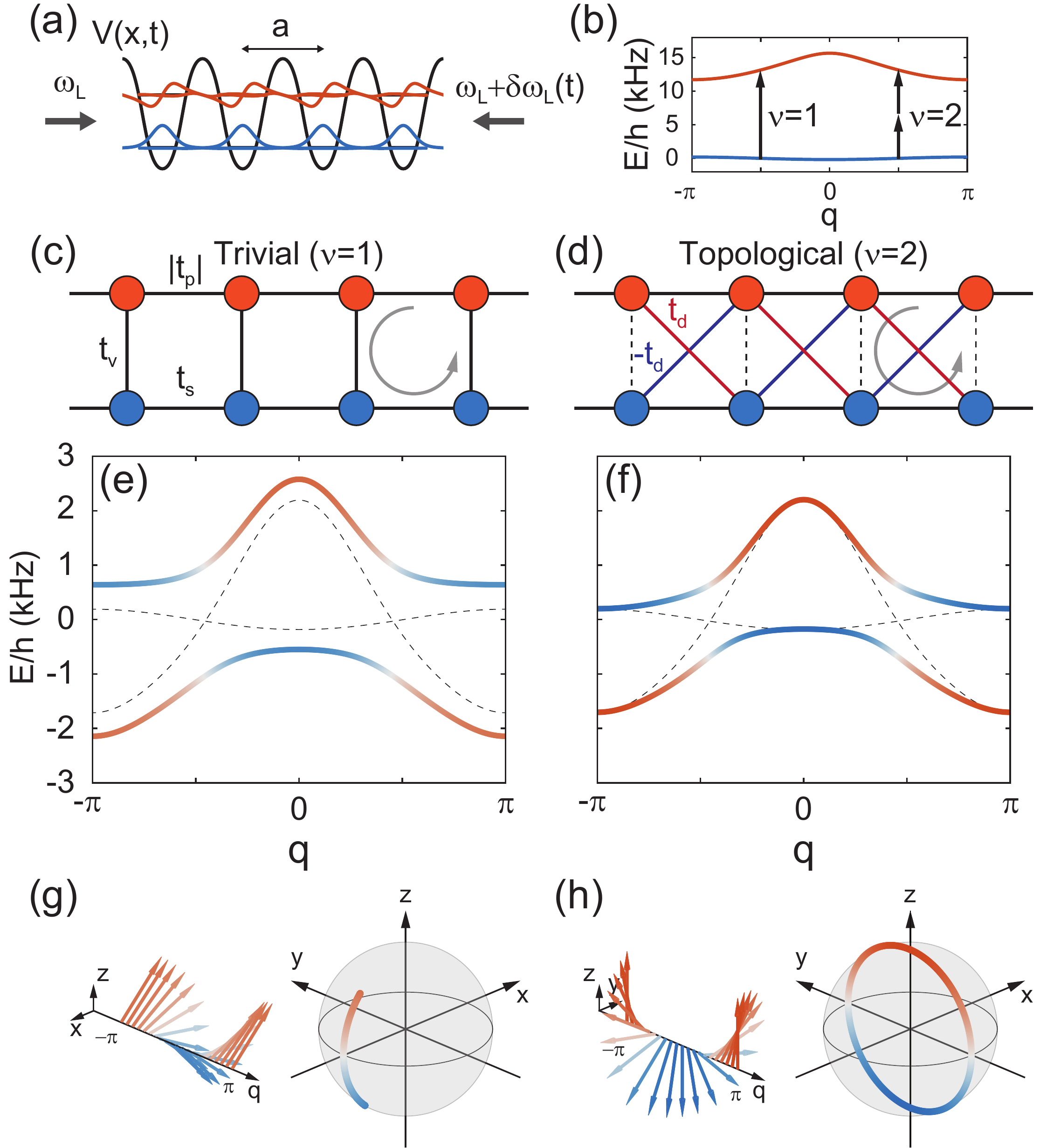}
\caption{ Topological Creutz ladder in a resonantly shaken optical lattice. (a) Schematic of the periodically driven 1D lattice system. The frequency difference between the two lattice laser beams is modulated as $\delta\omega (t) = A \sin(\omega t+\varphi)$, resulting in lattice vibrations with amplitude $d=\frac{A}{2\pi\omega}a$. (b) The $s$ and $p$ orbitals can be resonantly coupled via one-photon ($\nu$=1) or two-photon ($\nu$=2) processes. Two-leg ladder diagrams for the systems with (c) the $\nu$=1 and (d) $\nu$=2 resonant couplings. The gray arrow indicates a $\pi$ gauge flux piercing each ladder plaquette. Band structures of the ladder systems for our experimental conditions with (e) $\{\omega,A\}/2\pi = \{13.4, 4.0\}$~kHz and (f) $\{6.7, 8.0\}$~kHz, giving $t_v/h=1.0$~kHz and $t_d/h=0.4$~kHz, respectively. The dashed lines indicate the bare dispersion curves of the $s$ and $p$ bands for $\{t_s, t_p\}/h=\{0.1,-1.0\}$~kHz. (g),~(h) Corresponding pseudo-spin distributions of the ground bands and their trajectories on the Bloch sphere.}
\end{figure}


To describe our 1D shaken lattice system, we take a two-band tight-binding approximation, where the lattice system is regarded as a two-leg ladder system constituted by the $s$ and $p$ orbitals, and the inter-orbital coupling by lattice shaking is depicted as the inter-leg links between the two legs. To explicate the properties of the inter-leg links generated by the one- or two-photon inter-orbital resonant transitions, we obtain the effective Hamiltonian $H^{(\nu)}_{\text{eff}}$ of the system in a rotating frame with frequency $\nu\omega$ ($\nu=1,2$), respectively, using a high-frequency expansion method~\cite{Goldman14, Eckardt15, Bukov15,SI}. To lowest order, the effective Hamiltonian is expressed as $H^{(\nu)}_{\text{eff}}=H_0 + H_\nu$ with  
\begin{align}
H_0 = & \sum_j \bigg\{\Psi_j^\dagger (\Delta_\nu \sigma_z)\Psi_j -  \Big[\Psi_j^\dagger (\bar{t}_r\mathbb{I} +t_r\sigma_z)\Psi_{j+1}+\text{H.c.}\Big]\bigg\}   \nonumber \\
H_1 = &	\sum_j \Psi_j^\dagger \bigg( \frac{h_0^{sp}}{2}R(\varphi)\sigma_x\bigg)\Psi_j    \nonumber \\
H_2= &	\sum_j \bigg[ \Psi_j^\dagger  \bigg(\frac{h_0^{sp}h_1}{2\hbar\omega} R(2\varphi)i\sigma_y  \bigg)\Psi_{j+1}+\text{H.c.}\bigg],
\end{align}
where $\Psi_j^\dagger = (c_{j,p}^\dagger , c_{j,s}^\dagger)$, $c_{j,\alpha}$ is the annihilation operator for a spinless fermion in the Wannier state $|j,\alpha\rangle$ on lattice site $j$ in orbital $\alpha\in\{s,p\}$, and $\{\mathbb{I},\bm{\sigma}\}$ are the identity and Pauli matrices. $H_0$ includes the effective orbital energy term with $\Delta_\nu =(\epsilon_{sp}-\nu\hbar\omega)/2$ and the intra-leg tunneling term with $\bar{t}_r = \frac{t_p+t_s}{2}$, and $t_r =\frac{t_p-t_s}{2}$, where $t_\alpha$ is the nearest-neighbor (NN) tunneling amplitude of the $\alpha$ orbital. $H_\nu$ represents the dominant inter-orbital coupling generated by the $\nu$-photon processes, where $h_\ell^{\alpha\beta}$$=\hbar A \frac{a}{2\pi} \langle j,\alpha|\frac{\partial}{\partial x} |j$$+$$\ell,\beta\rangle$, $h_1=h_1^{pp}-h_1^{ss}$, and $R(\varphi) = \text{exp}(-i\varphi\sigma_z)$~\cite{SI}.

The two-leg ladder system is characterized by the inter-leg link configuration set by $H_\nu$. $H_1$ describes the on-site orbital-changing transitions, corresponding to the direct inter-leg links with amplitude $t_v$$=$$h_{0}^{sp}/2$. On the other hand, $H_2$ describes the second-order processes consisting of on-site orbital changing and NN tunneling, which amount to the diagonal inter-leg links with amplitude $t_d$$=$$h_0^{sp}h_1/(2\hbar\omega)$. Since the lattice shaking is an odd-parity operation, the two-photon on-site transition between the $s$ and $p$ orbitals with opposite parities is forbidden by parity conservation~\cite{Zhang14}. The resulting ladder systems for $\nu=1$ and 2 are schematically described in Figs.~1(c) and (d), respectively. The gray circulating arrow indicates a $\pi$ gauge flux piercing each ladder plaquette, accounting for the opposite signs of $t_p$ and $t_s$. Thus, under the two-photon resonance condition, the shaken lattice system realizes the seminal Creutz ladder model~\cite{Creutz99}. The cross-linked two-leg ladder for spinless fermions is equivalent to a spin-1/2 1D chain with spin-flip hopping~\cite{Song18, Liu13}.

 The Bloch Hamiltonian of the two-leg ladder system is expressed as $H^{(\nu)}_q=-2\bar{t}_r \cos(q)+ \bm{B}_\nu(q)\cdot\bm{\sigma}$, where $q$ is the quasimomentum in units of $a^{-1}$ and the effective magnetic field $\bm{B}_\nu(q)$ is given by 
\begin{align}
\bm{B}_1(q)=& 	t_v\hat{\bm{\rho}}_1 + [\Delta_1 - 2t_r \cos(q)]\hat{\bm{z}}  \nonumber \\
\bm{B}_2(q)=&	2t_d \sin(q) \hat{\bm{\rho}}_2 + [\Delta_2 - 2t_r \cos(q)]\hat{\bm{z}}
\end{align}
with $\hat{\bm{\rho}}_1=\cos(\varphi) \hat{\bm{x}}+ \sin(\varphi) \hat{\bm{y}}$ and $\hat{\bm{\rho}}_2=-\sin(2\varphi) \hat{\bm{x}}+ \cos (2\varphi) \hat{\bm{y}}$. The band structures of the ladder system with $\Delta_\nu=0$ for $\nu=1$ and 2 are displayed in Figs.~1(e) and 1(f), respectively, and the corresponding pseudo-spin distributions of the ground bands and their trajectories on the Bloch sphere as $q$ changes over its space are shown in Figs.~1(g) and (h). The half of the solid angle subtended by the trajectory is the geometric representation of the Zak phase $\gamma_Z$ of the band~\cite{Zak89}. We have $\gamma_Z=\pi$ for $\nu=2$, whereas $\gamma_Z=0$ for $\nu=1$. Having non-zero $\gamma_Z$ is the key topological character of the cross-linked Creutz ladder system~\cite{Creutz99,Hugel14}, which is preserved once $\bm{B}_2(q)$ has a winding structure with $|\Delta_2|<2|t_r|$. It is noted that  one of the topological bands becomes dispersionless when $|\Delta_2|=2|\bar{t}_r|$ and $t_d^2=|t_s t_p|$. In this work, our focus is to experimentally demonstrate the winding of $\bm{B}_2(q)$ in the resonantly shaken lattice system.


We first investigate the resonance condition for inter-obital coupling by measuring the momentum distribution $n(k)$ of the atoms as a function of the driving frequency $\omega$. After preparing the atoms in the $s$ band of the static lattice, we suddenly applied the periodic driving over 20 cycles, and measure $n(k)$ using a band mapping technique~\cite{Kang18}. In Fig.~2, the change in the momentum distribution, $\Delta n(k)$, from that of the non-driven sample is displayed as a function of $\omega$, where $n(k)$ is normalized as $\int n(k)dk=2\pi$ and $k$ is expressed in units of $a^{-1}$. A strong spectral signal was observed in the range $\omega/2\pi$$=11$$\sim$$17~\text{kHz}$~[Fig.~2(a)]. The spectral peak position follows the one-photon resonance condition $\omega=\omega_{sp}(q) \approx[\epsilon_{sp}-4t_r \cos (q)]/\hbar$, where $\{\epsilon_{sp},t_r\}/h=\{13.4,-0.54\}$~kHz for our lattice parameters. The two-photon $s$-$p$ coupling was observed in the corresponding half-frequency range $\omega/2\pi$$=$$5.5$$\sim$$8~\text{kHz}$~[Fig.~2(b)]. The spectral structure appears consistent with the resonance condition of $2\omega=\omega_{sp}(q)$, but compared with the one-photon resonance case, the signal strength is significantly suppressed at $k=0$ and $\pm \pi$ [Figs.~2(c) and (d)]. This is attributed to the $q$-dependence of the transverse field amplitude, $B_{2,\rho}(q)\propto \sin(q)$, which determines the coupling strength between the two orbital-momentum states, $|q,s\rangle$ and $|q,p\rangle$. The two-photon and three-photon couplings between the $s$ and $d$ orbitals were also observed for $\omega/2\pi=11$$\sim$14~kHz and $\omega/2\pi\approx 8.5~\text{kHz}$, respectively, with substantial population transfer to the high-momentum region of $|k|>2\pi$.

\begin{figure}[t]
\includegraphics[width=8.0cm]{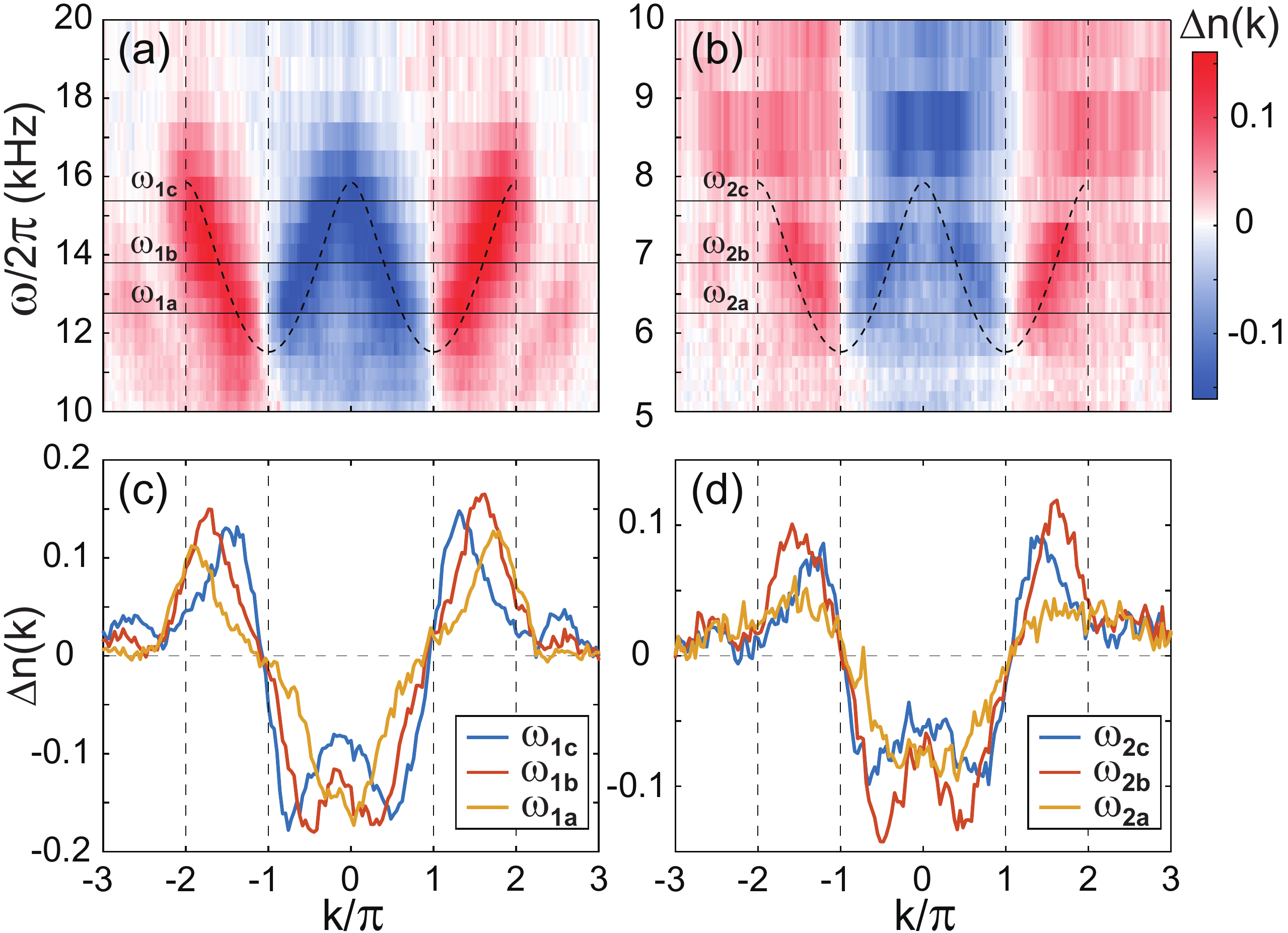}
\caption{Shaking spectroscopy of fermionic atoms in an optical lattice. (a),~(b) Spectra of the momentum distribution change $\Delta n(k)$ as a function of the driving frequency $\omega$.  The driving amplitude $A/2\pi=2$~kHz in (a) and 4~kHz in (b). The dashed lines in (a) and (b) indicate the one-photon and two-photon resonance conditions, i.e., $\omega=\omega_{sp}(q)$ and $2\omega=\omega_{sp}(q)$, respectively~\cite{footnote1}. (c),~(d) Spectral profiles of $\Delta n(k)$ at various $\omega$, indicated by the horizontal lines in (a) and (b).}
\end{figure}


\begin{figure}
	\includegraphics[width=8.2cm]{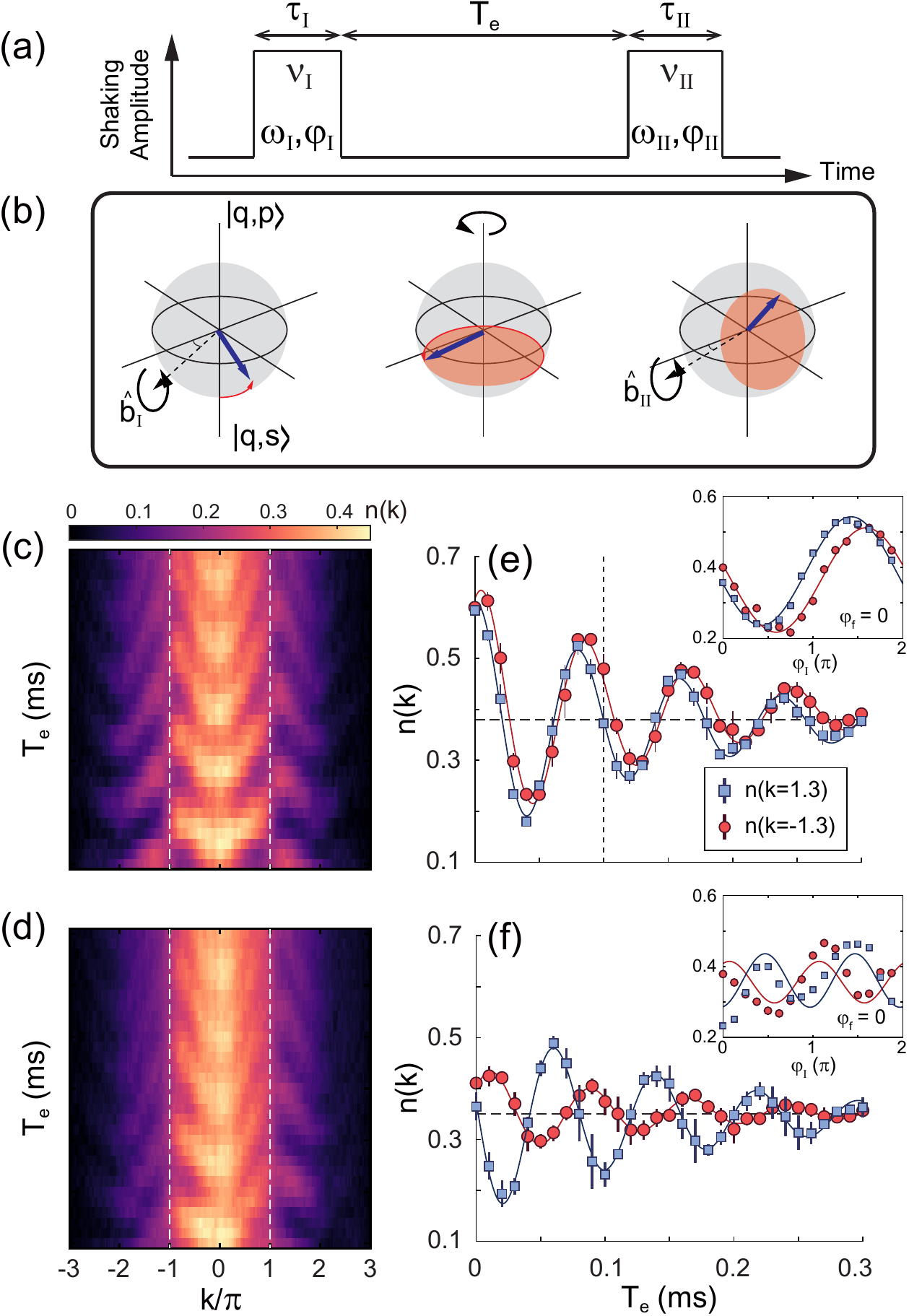}
	\caption{Momentum-resolved Ramsey interferometry using two separate pulses of lattice shaking. Schematic of (a) the lattice shaking sequence and (b) the pseudo-spin evolution on the Bloch sphere for the quasimomentum $q$ state. $\hat{\bm{b}}_\text{I(II)}$ indicates the effective magnetic field direction for the first (second) shaking pulse. The orange disk denotes the precession plane of the pseudo-spin during the intermittent period. Momentum distribution $n(k)$ as a function of $T_e$ for (c) $(\nu_\text{I},\nu_\text{II})=(1,1)$ and (d) (2,1), where $\varphi_\text{I(II)}=0$ and $\tau_\text{I(II)}=2\pi/\omega_\text{I(II)}$.  (e),~(f) $n(k)$ at $k = \pm 1.3\pi$ as a function of $T_e$ in (c) and (d), respectively. The solid lines are the damped sinusoidal function fits to the data. The inset shows $n(k$$=$$\pm 1.3\pi)$ as a function of the modulation phase $\varphi_\text{I}$ of the first shaking for $T_e=100~\mu$s. Each data point was obtained from seven measurements and its error bar indicates their standard deviation.}
\end{figure}

To further examine the structure of $\bm{B}_\nu(q)$ in the $q$ space, we employed an interferometric measurement technique using two pulses of resonant lattice shaking separated by a time interval $T_e$ [Fig.~3(a)]. The frequency modulations were set as $\delta \omega_L(t)=A_{\text{I}} \sin (\omega_\text{I} t +\varphi_\text{I})$ for the first pulse in $0<t<\tau_\text{I}$ and $\delta \omega_L(t)=A_{\text{II}} \sin (\omega_\text{II} t'+\varphi_\text{II})$ with $t'=t-(\tau_\text{I}+T_e)$ for the second pulse in $0<t'<\tau_\text{II}$. Here, $\hbar\omega_\text{I(II)}=\epsilon_{sp}/\nu_\text{I(II)}$ with $\nu_\text{I(II)}$=1 or 2, and $\tau_\text{I(II)}$ is the first (second) pulse duration. The measurement scheme can be described as Ramsey interferometry, where two pulses of lattice shaking play the roles of rotating magnetic fields [Fig.~3(b)]. In the rotating frame with frequency $\epsilon_{sp}/\hbar$, an atom was initially prepared in the $|q,s\rangle$ state in the presence of the axial magnetic field of $B_{z}(q)=2t_r\cos(q)$ and it experienced the transverse fields of $B_{\nu_\text{I(II)},\rho}(q)$ along the $\hat{\bm{\rho}}_{\nu_\text{I(II)}}$ directions over the pulse durations. Then, the interferometric signal was obtained by measuring the atom population of the $|q,p\rangle$ state.

Because the Ramsey fringe signal is determined by the directions of the pulsed transverse fields and the accumulated precession angle, its $q$-dependence faithfully reflects the structure of the effective magnetic field. For example, in the case of $(\nu_\text{I},\nu_\text{II})$=$(1,1)$, the transverse field is $B_{1,\rho}$$=t_v$, uniform over the whole $q$ space, and the Ramsey signal will appear as a conventional form of $S(q)\sim \frac{1}{2}[1+\cos\{\omega_{sp}(q)T_e+\varphi_\text{I}-\varphi_\text{II}\}]$~\cite{SI}. Here, the evolution rate with $T_e$ directly reveals $B_z(q)$ via $\omega_{sp}(q)$. Another interesting case is the one with $(\nu_\text{I},\nu_\text{II})$=$(2,1)$, where the first transverse field has opposite directions for $q>0$ and $q<0$, i.e., $B_{2,\rho}(-q)=-B_{2,\rho}(q)$, so the Ramsey signal would be asymmetric in $q$ as $S(q)\sim\frac{1}{2}[1-\text{sgn}(q)\times\sin\{\omega_{sp}(q)T_e+2\varphi_\text{I}-\varphi_\text{II}\}]$~\cite{SI}. As $B_{2,z}=2t_r \cos(q)$ was confirmed by the two-photon resonance condition of $2\omega=\omega_{sp}(q)$, the establishment of the asymmetric relation of $B_{2,\rho}(-q)=-B_{2,\rho}(q)$ would be sufficient for verifying the winding structure of $\bm{B}_2$.

In Figs.~3(c) and 3(d), we display the measurement results of the Ramsey interferometry for $(\nu_\text{I},\nu_\text{II})=(1,1)$ and $(\nu_\text{I},\nu_\text{II})=(2,1)$, respectively, as a function of $T_e$. Here $(\varphi_\text{I},\varphi_\text{II})=(0,0)$ and $\tau_\text{I(II)}= 2\pi/\omega_\text{I(II)}$. We set $A/2\pi = 5$ or 8~kHz for $\nu_\text{I(II)}=1$ or 2, respectively, to obtain a $\pi/2$-pulse for $q=\pm 0.7\pi$. The population oscillations are clearly observed for each $q$ and the oscillation frequency with $T_e$ is found to be in good quantitative agreement with $\omega_{sp}(q)$~\cite{SI}. Also, it was observed that the Ramsey signals are mirror-symmetric with respect to $k=0$ for $\nu_\text{I}=1$ and asymmetric for $\nu_\text{I}=2$ as expected. In Figs.~3(e) and 3(f), we display the population evolutions at $k=\pm1.3\pi$ with increasing $T_e$ for $\nu_\text{I}=1$ and 2, respectively. From the damped sinusoidal functions fit to the data, we measure the phase difference between the two oscillation curves to be $0.10 \pi$ for $\nu_\text{I}=1$ and $0.77 \pi$ for $\nu_\text{I}=2$, which are slightly different from the expected values of 0 and $\pi$, respectively.  Such a deviation might be attributed to the off-resonant coupling effect~\cite{SI}, which was neglected in our description. The dependence of the Ramsey signal on the driving phases $\varphi_\text{I}$ and $\varphi_\text{II}$ was also examined and confirmed to be consistent with our description~\cite{SI}. In particular, the double periodicity of the Ramsey signal with increasing $\varphi_\text{I}$ was observed  in the $\nu_\text{I}=2$ case~[Fig.~3(f) inset], which is a direct consequence of the fact that two photons are involved in the inter-orbital coupling.

\begin{figure}
	\includegraphics[width=8.4cm]{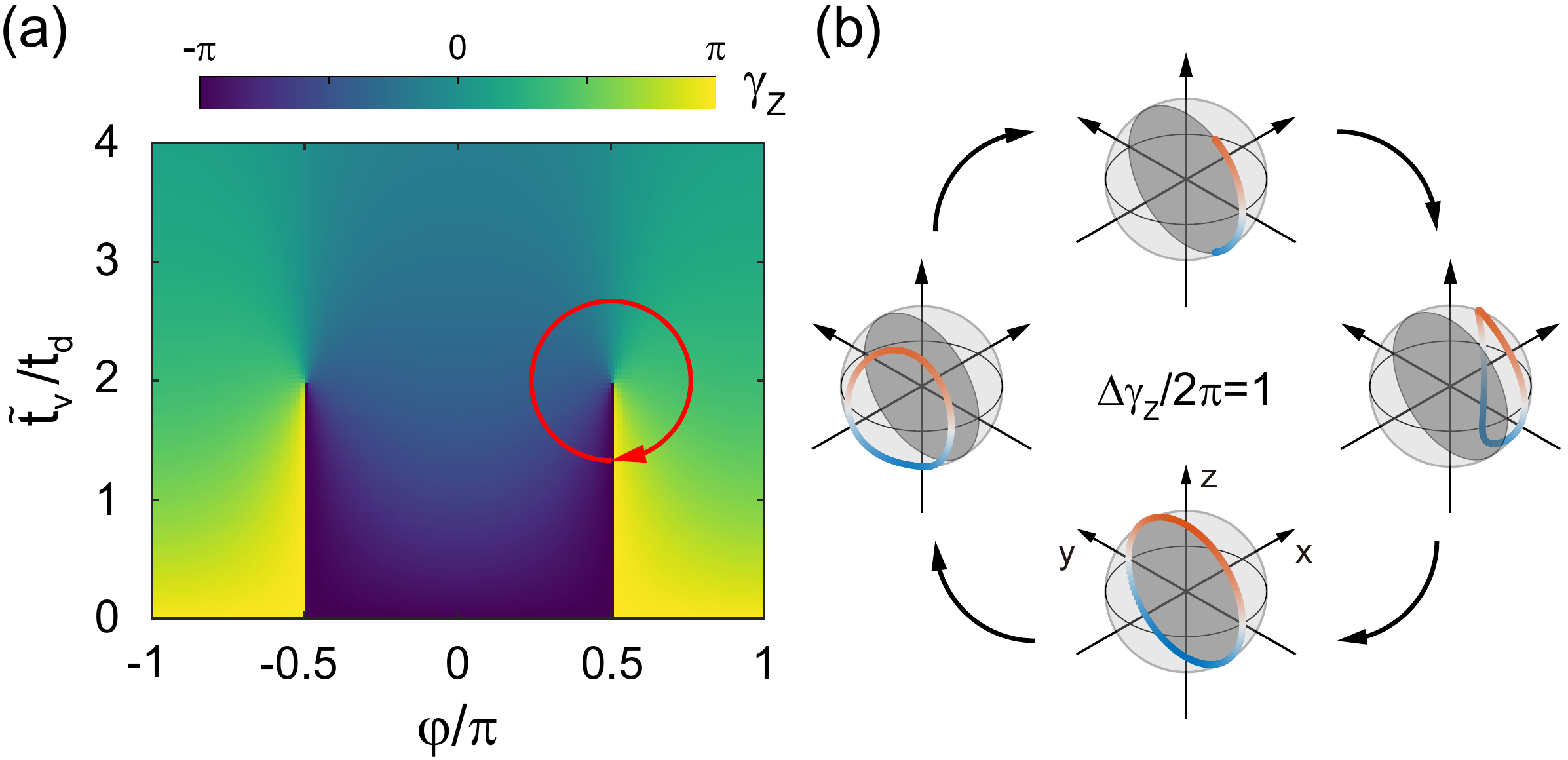}
	\caption{Topological charge pumping in the Creutz ladder. (a) Zak phase $\gamma_Z$  in the plane of $\tilde{t}_v/t_d$ and $\varphi$. The critical points are located at $\{\tilde{t}_v/t_d,\varphi\}=\{2,\pm\pi/2\}$. Along the path encircling one of the critical points, $\gamma_Z$ continuously changes by $2\pi$.  (b) Evolution cycle of the pseudo-spin trajectory of the ground band on the Bloch sphere as the system adiabatically moves along the red encircling path in (a).}
\end{figure}


The observation that one- and two-photon resonant couplings bring about the distinctive effects of direct and cross inter-leg links, respectively, prompted us to discuss the extension of our Creutz ladder scheme using both types of resonant couplings. Specifically, we considered a situation in which the lattice shaking was applied as $\delta\omega_L(t) =A \sin(\omega t)+ \tilde{A} \sin (2\omega t+\varphi)$ with $\Delta_2=0$. Here, the additional $2\omega$ driving provides  one-photon resonant coupling. In our effective Hamiltonian description, we found that the effects of the two drivings were additive~\cite{SI}, resulting in the effective magnetic field 
\begin{align}
\bm{B}(q)=\tilde{t}_v \hat{\bm{\rho}}_1(\varphi) +2t_d \sin(q) \hat{\bm{y}} + -2t_r \cos(q) \hat{\bm{z}}, \nonumber
\end{align}
with $\tilde{t}_v=(\tilde{A}/A)t_v$. This shows that the direct inter-leg link of the Creutz ladder is generated by the additional $2\omega$ driving; furthermore, its complex amplitude can be directly controlled with the driving parameters $\tilde{A}$ and $\varphi$.

The flexible control of the inter-leg links with two-tone driving would enable to study the topological phase transitions of the Creutz ladder system, which occurs at $\{\tilde{t}_v,\varphi\}_c=\{2t_d,\pm \pi/2\}$, accompanied by the band gap closing for $\bm{B}(q)=0$ at $q=\mp \pi/2$ and a sudden change of the Zak phase $\gamma_Z$ by $\pi$. Interestingly, a topological charge pump can be realized by dynamically controlling the system  parameters of $\{\tilde{t}_v,\varphi\}$ to encircle one of the critical points (Fig.~4)~\cite{Sun17}. As the system adiabatically moves along the closed pumping path, $\gamma_Z$ continuously changes by $2\pi$ per cycle, which would result in one lattice site shift of atoms in the ladder. In our experiment with $\Delta_2=0$ and $A/2\pi=8$~kHz ($t_d/h=0.4$~kHz), the atom loss rate out of the $s$-$p$ ladder system was observed to be $\approx 0.17 t_d$. For robust implementation of the pump scheme, further experimental optimization seems to be necessary such as phase stabilization of the lattice laser beams and transverse confinement of atoms with additional $yz$ lattices beams.

In conclusion, we demonstrated the realization of the topological Creutz ladder in a periodically shaken 1D optical lattice via two-photon resonant coupling. Besides the implementation of the topological charge pump using the two-tone driving, we may pursue topological flat band engineering with our ladder system, which would be interesting for possible emergence of correlated topological phases such as fractional quantum Hall states~\cite{Neupert11,Bergholtz13}. Also, we expect that the two-photon resonant coupling method can be readily applied to 2D optical lattice systems, providing an alternative route to investigate anomalous quantum Hall states~\cite{Goldman13,Liu10}.

We thank Wei Zheng and Hui Zhai for helpful discussions. This work was supported by the Institute for Basic Science in Korea (Grant No. IBS-R009-D1) and the National Research Foundation of Korea (Grants No. NRF-2018R1A2B3003373, and No. 2014-H1A8A1021987).


\clearpage
\newpage 

\setcounter{equation}{0}
\setcounter{figure}{0}
\renewcommand{\thefigure}{S\arabic{figure}}
\renewcommand{\theequation}{S\arabic{equation}}

\section{Supplemental Material}

\subsection{Atom in a driven optical lattice}
In the experiment, a 1D optical lattice was formed by spatially interfering two laser beams. The lattice potential is expressed as 
\begin{equation}
V(x) = \frac{V_L}{2} \cos\Big[\frac{2\pi}{a}\big(x-x_0\big)\Big], \nonumber
\end{equation}
where $a$ is the lattice spacing and $x_0=\frac{a}{2\pi} \phi$ represents the lattice position determined by the relative phase $\phi$ between the two laser beams. Periodic shaking of the lattice was achieved by sinusoidally modulating the frequency difference between the two lattice laser beams as $\delta\omega_L(t) = A\sin(\omega t+\varphi)$ with the modulation frequency $\omega$, the amplitude $A$, and the phase $\varphi$. From the relation,  $\phi(t)-\phi(0) =\int_{0}^t \delta\omega_L(\tau)d\tau$, the lattice position is given by $x_0(t)=-d\cos(\omega t + \varphi)$ with $d =\frac{A}{2\pi\omega}a$. 

The Hamiltonian of an atom in the periodically driven optical lattice is given by
\begin{equation}
H_{Lab} = \frac{p^2}{2m} + \frac{V_L}{2}\cos\Big[ \frac{2\pi}{a} \big(x + d\cos(\omega t+\varphi)\big) \Big].
\end{equation}
Taking the unitary transformation of  
\begin{equation}
U_1(t) =\exp\Big[ \frac{i}{\hbar} d \cos(\omega t+\varphi) p\Big]
\end{equation}
for  $x \rightarrow x-d\cos(\omega t+\varphi)$, we obtain the system's Hamiltonian $H$ in the reference frame comoving with the optical lattice, and $H$ is expressed as $H=H_\text{stat}+\delta H$ with 
\begin{align}
H_\text{stat} &= \frac{p^2}{2m}+\frac{V_L}{2}\cos\big(\frac{2\pi}{a} x\big) \nonumber \\
\delta H & = -d\omega \sin(\omega t+\varphi)p,
\end{align}
where $H_\text{stat}$ is the Hamiltonian of the stationary lattice system and $\delta H$ represents the perturbation from the inertial force induced by the lattice shaking.

\subsection{Two-band model description}

In the two-band tight binding approximation, the Hamiltonian of the lattice system is expressed as
\begin{equation}
H = \sum_j \Psi_j^\dagger K(t) \Psi_j -\sum_j [\Psi_j^\dagger J(t) \Psi_{j+1}+\text{H.c.}],
\end{equation}
where $\Psi_j^\dagger = ( c_{j,p}^\dagger,c_{j,s}^\dagger)$, $c_{j,\alpha}$ is the annihilation operator for the atom in the Wannier state $|j,\alpha\rangle$ on lattice site $j$ in orbital $\alpha\in\{s,p\}$. The matrices $K(t)$ and $J(t)$ are given by
\begin{align}
K(t)  &= 
	\begin{pmatrix}
	\epsilon_p & -ih_0^{sp}\sin(\omega t+\varphi) \\
	ih_0^{sp}\sin(\omega t+\varphi) & \epsilon_s
	\end{pmatrix}  \nonumber \\
J(t) &=
	\begin{pmatrix}
	t_p-ih_1^{pp}\sin(\omega t+\varphi) & ih_1^{sp}\sin(\omega t+\varphi) \\
	-ih_1^{sp}\sin(\omega t+\varphi) & t_s-ih_1^{ss}\sin(\omega t+\varphi)
	\end{pmatrix} \nonumber
\end{align}
with
\begin{align}
\epsilon_\alpha =& \langle j, \alpha|H_\text{stat} |j,\alpha\rangle \nonumber \\
t_\alpha =& -\langle j, \alpha| H_\text{stat}|j+1,\alpha\rangle \nonumber \\
h_\ell^{\alpha\beta} =&~\hbar A \frac{a}{2\pi} \langle j, \alpha|(\partial/\partial x)|j+\ell,\beta\rangle. 
\end{align}
 $\epsilon_\alpha$ and $t_\alpha$ are the on-site energy and nearest-neighbor hopping amplitude of the $\alpha$ orbital, respectively, and $h_\ell^{\alpha\beta}$ is the transition element between the $\alpha$ and $\beta$ orbitals separated by $\ell$ lattice sites. In terms of the identity and Pauli matrices $\{ \mathbb{I},\sigma_x,\sigma_y,\sigma_z\}$, 
\begin{align}
K(t) =&~ \bar{\epsilon} \mathbb{I}+ \frac{\epsilon_{sp}}{2} \sigma_z +h_0^{sp}\sin(\omega t+\varphi)\sigma_y \nonumber \\
J(t) =&~[\bar{t}_r-i\bar{h}_1 \sin(\omega t+\varphi)] \mathbb{I} \nonumber \\
&+ [t_r-ih_1 \sin(\omega t+\varphi)] \sigma_z \nonumber \\
&-h_1^{sp}\sin(\omega t+\varphi) \sigma_y	
\end{align}
with $\bar{\epsilon}=\frac{\epsilon_p+\epsilon_s}{2}$, $\epsilon_{sp}=\epsilon_p-\epsilon_s$, $\bar{t}_r =\frac{t_p+ t_s}{2}$, $t_r =\frac{t_p-t_s}{2}$, $\bar{h}_1 = \frac{h_1^{pp}+h_1^{ss}}{2}$, and $h_1 = \frac{h_1^{pp}-h_1^{ss}}{2}$.

To investigate the effects of the $\nu$-photon inter-orbital resonant coupling, we obtain the Hamiltonian $H'$ in a rotating reference frame with frequency $\nu \omega$, taking a unitary transformation of
\begin{equation}
U_2(t) = \sum_{j}\Big[ e^{-i\frac{\nu \omega}{2} t}c_{j,p}^\dagger c_{j,p} +e^{i\frac{\nu\omega}{2} t}c_{j,s}^\dagger c_{j,s}\Big].
\end{equation}
 The two matrices $K(t)$ and $J(t)$ are transformed to, respectively,
\begin{align}
K'(t) =&~\Delta_\nu \sigma_z +h_0^{sp}\sin(\omega t+\varphi)R(-\nu\omega t)\sigma_y \nonumber \\
J'(t) =&~[\bar{t}_r-i\bar{h}_1 \sin(\omega t+\varphi)] \mathbb{I} \nonumber \\
&+ [t_r-ih_1 \sin(\omega t+\varphi)] \sigma_z \nonumber \\
	&-h_1^{sp}\sin(\omega t+\varphi)R(-\nu\omega t) \sigma_y,	
\end{align}
where $\Delta_\nu = (\epsilon_{sp}-\nu\hbar\omega)/2$ and $R(\theta) =\text{exp}(-i\theta\sigma_z)$. Here we ignore the energy offset of $\bar{\epsilon}\mathbb{I}$ in $K(t)$.

\subsubsection{Effective Hamiltonian for $\nu=1$}
Using a high frequency expansion method up to second order processes~\cite{Eckardt15}, the effective Floquet Hamiltonian of the periodically driven system is obtained as
\begin{equation}
H_{\text{eff}}^{(\nu)} = H_0 + \sum_{n=1}^\infty \frac{[H_n,H_{-n}]}{n\hbar\omega},
\end{equation}  
where $H_n$ is the $n\omega$ Fourier component of the original time-periodic Hamiltonian $H'(t)$, such that
\begin{equation}
H'(t) = H_0 +\sum_{n\neq 0} H_n e^{i n \omega t}.
\end{equation}

In the case of $\nu=1$, from Eq.~(S8),
\begin{align}
H_0 &= \sum_j \Psi_j^\dagger \bigg[\Delta_1\sigma_z+\frac{h_0^{sp}}{2}R(\varphi)\sigma_x\bigg]\Psi_j    \nonumber \\
	   &+\sum_j \bigg\{ \Psi_j^\dagger\bigg[-\bar{t}_r \mathbb{I} -t_r \sigma_z+\frac{h_1^{sp}}{2}R(\varphi)\sigma_x\bigg]\Psi_{j+1}+\text{H.c.}\bigg\} \nonumber \\
H_1 &=\sum_{j} \bigg\{ \Psi_j^\dagger\bigg[\frac{e^{i\varphi}}{2}(\bar{h}_1\mathbb{I}+h_1\sigma_z) \bigg]\Psi_{j+1} \nonumber \\
	&~~~~~~~~~~+\Psi_{j+1}^\dagger\bigg[-\frac{e^{i\varphi}}{2}(\bar{h}_1\mathbb{I}+h_1\sigma_z) \bigg]\Psi_{j}\bigg\} \nonumber \\
H_2 &= \sum_j \Psi_j^\dagger \bigg[-\frac{h_0^{sp}}{2}e^{i\varphi}\sigma^+ \bigg]\Psi_j \nonumber \\
&+\sum_{j} \bigg\{ \Psi_j^\dagger\bigg[\frac{-h_1^{sp}}{2}e^{i\varphi}\sigma^+ \bigg]\Psi_{j+1}+\Psi_{j+1}^\dagger\bigg[\frac{-h_1^{sp}}{2}e^{i\varphi}\sigma^+ \bigg]\Psi_{j}\bigg\}
\end{align}
and $H_{-n} = H_n^\dagger$, where $\sigma^\pm = (\sigma_x \pm i\sigma_y)/2$. Then,
\begin{align}
H_{\text{eff}}^{(1)} &= \nonumber \\
\sum_j &\Psi_j^\dagger \bigg[\bigg(\Delta_1+\frac{(h_0^{sp})^2+(h_1^{sp})^2}{8\hbar\omega}\bigg)\sigma_z+\frac{h_0^{sp}}{2}R(\varphi)\sigma_x\bigg]\Psi_j    \nonumber \\
	   +\sum_j& \bigg\{ \Psi_j^\dagger\bigg[-\bar{t}_r \mathbb{I}+\bigg(\frac{h_0^{sp}h_1^{sp}}{4\hbar\omega}-t_r\bigg) \sigma_z+\frac{h_1^{sp}}{2}R(\varphi)\sigma_x\bigg]\Psi_{j+1}\nonumber \\
	   &~~+\Psi_j^\dagger \bigg[ \frac{(h_1^{sp})^2}{8\hbar\omega} \sigma_z \bigg]\Psi_{j+2}+ \text{H.c.}\bigg\}. 
\end{align}
When $\hbar\omega\gg h_0^{sp} \gg |h_1^{sp}|$, the effective Hamiltonian is approximated as
\begin{align}
H_{\text{eff}}^{(1)} &\approx  \sum_j \Psi_j^\dagger \bigg[\Delta_1\sigma_z+t_v R(\varphi)\sigma_x\bigg]\Psi_j    \nonumber \\
	   &-\sum_j \bigg\{ \Psi_j^\dagger (\bar{t}_r \mathbb{I}+t_r \sigma_z)\Psi_{j+1}+\text{H.c.}\bigg\}.
\end{align}
with $t_v=h^{sp}_0/2$. The Bloch Hamiltonian is given by 
\begin{equation}
H_q^{(1)} =-2\bar{t}_r\cos(q)\mathbb{I}+\bm{B}_1(q) \cdot \bm{\sigma} \nonumber
\end{equation}
with
\begin{align}
\bm{B}_1(q) =&~ t_v \hat{\bm{\rho}}_1+ [\Delta_1-2t_r\cos(q)]\hat{\bm{z}} \nonumber \\
\hat{\bm{\rho}}_1=&\cos(\varphi)\hat{\bm{x}}+\sin(\varphi)\hat{\bm{y}}, \nonumber \\
\end{align}
where $q$ is the quasimomentum normalized in units of $a^{-1}$.

\subsubsection{Effective Hamiltonian for $\nu=2$}
In the case of $\nu =2$,
\begin{align}
H_0 = \sum_j & \Psi_j^\dagger (\Delta_2\sigma_z)\Psi_j -\sum_j \bigg[ \Psi_j^\dagger(\bar{t}_r\mathbb{I} +t_r\sigma_z)\Psi_{j+1}+\text{H.c.}\bigg] \nonumber \\
H_1 =\sum_{j} & \Psi_j^\dagger \bigg[ \frac{h_0^{sp}}{2}e^{-i\varphi}\sigma^+ \bigg]\Psi_{j}   \nonumber \\
	   +\sum_{j} &\bigg\{ \Psi_j^\dagger \bigg[\frac{h_1^{sp}}{2}e^{-i\varphi}\sigma^+ +\frac{e^{i\varphi}}{2}(\bar{h}_1\mathbb{I}+h_1\sigma_z) \bigg]\Psi_{j+1} \nonumber\\
	   &+\Psi_{j+1}^\dagger \bigg[\frac{h_1^{sp}}{2}e^{-i\varphi}\sigma^+ -\frac{e^{i\varphi}}{2}(\bar{h}_1\mathbb{I}+h_1\sigma_z) \bigg] \Psi_{j}\bigg\} \nonumber \\
H_3 = \sum_j & \Psi_j^\dagger \bigg[-\frac{h_0^{sp}}{2}e^{i\varphi}\sigma^+ \bigg]\Psi_j \nonumber \\
+\sum_{j}& \bigg\{ \Psi_j^\dagger\bigg[\frac{-h_1^{sp}}{2}e^{i\varphi}\sigma^+ \bigg]\Psi_{j+1}+\Psi_{j+1}^\dagger\bigg[\frac{-h_1^{sp}}{2}e^{i\varphi}\sigma^+ \bigg]\Psi_{j}\bigg\},
\end{align}
and $H_{-n} = H_n^\dagger$, giving 
\begin{align}
H_{\text{eff}}^{(2)}&= \nonumber\\
 \sum_j&  \Psi_j^\dagger \bigg[\bigg(\Delta_2+\frac{(h_0^{sp})^2+(h_1^{sp})^2}{3\hbar\omega}\bigg)\sigma_z -\frac{h_1^{sp}h_{1}}{2\hbar\omega}R(2\varphi) \sigma_x\bigg]\Psi_j \nonumber \\
+\sum_j &\bigg\{ \Psi_j^\dagger \bigg[-\bar{t}_r\mathbb{I} +\bigg(\frac{2h_0^{sp}h_1^{sp}}{3\hbar\omega}-t_r \bigg)\sigma_z  +\frac{h_0^{sp}h_1}{2\hbar\omega}R(2\varphi)i\sigma_y\bigg]\Psi_{j+1}   \nonumber \\
	   & +\Psi_j^\dagger\bigg[\frac{(h_1^{sp})^2}{3\hbar\omega} \sigma_z +\frac{h_1^{sp}h_{1}}{2\hbar\omega}R(2\varphi) i \sigma_y \bigg]\Psi_{j+2} +\text{H.c.}\bigg\}. 
\end{align}
With $\hbar\omega \gg h^{sp}_0 > |h_1| > |h^{sp}_1|$, $H^{(2)}_\text{eff}$ is approximated as
\begin{align}
H_{\text{eff}}^{(2)} &\approx   \sum_j  \Psi_j^\dagger (\Delta_2\sigma_z)\Psi_j \nonumber \\
&-\sum_j  \bigg\{ \Psi_j^\dagger \bigg[ \bar{t}_r\mathbb{I} +t_r\sigma_z - t_d R(2\varphi)i\sigma_y\bigg]\Psi_{j+1}+\text{H.c.}\bigg\}
\end{align}
with $t_d=h_0^{sp} h_1 / (2\hbar \omega)$. Here, we ignore the ac Stark shift, $\delta_2^{ac}\approx \frac{(h_0^{sp})^2}{3\hbar\omega}$, in the orbital energy, which is comparable to $t_d$ in magnitude but much smaller than $t_r$ in the experiment. The Bloch Hamiltonian is given by
\begin{equation}
H_q^{(2)} =-2\bar{t}_r \cos(q)\mathbb{I}+\bm{B}_2(q) \cdot \bm{\sigma} \nonumber
\end{equation}
with
\begin{align}
\bm{B}_2(q) =&~ 2t_d \sin(q) \hat{\bm{\rho}}_2+[\Delta_2- 2t_r\cos(q)] \hat{\bm{z}}\nonumber\\
\hat{\bm{\rho}}_2=&-\sin(2\varphi)\hat{\bm{x}}+\cos(2\varphi)\hat{\bm{y}}.
\end{align} 

For our experimental condition, $h_0^{sp}:|h_1|:|h_1^{sp}|=6.4:1.7:0.3$ and $h_0^{sp}/h<1.3$~kHz, whereas $\omega/2\pi=13.7$ or 6.8~kHz, and $|t_r|/h=0.54$~kHz, justifying the approximation taken in Eqs.~(S13) and (S17).

\subsection{Ramsey interferometry using two pulses of lattice shaking}

We performed the Ramsey interferometry by modulating the frequency difference $\delta\omega_L$ as
\begin{align}
\delta\omega_L (t)= 
	\begin{dcases*}
		A_{\text{I}} \sin (\omega_\text{I} t +\varphi_\text{I})    &  for $0<t<\tau_\text{I}$, \\
		~~~~~~0 & for $\tau_\text{I}<t<\tau_\text{I}+T_e$, \\
		A_{\text{II}} \sin (\omega_\text{II} t'+\varphi_\text{II}) & for $0<t'<\tau_\text{II}$
	\end{dcases*} \nonumber
\end{align}
with $t'=t-(\tau_\text{I}+T_e)$. Here, $\omega_\text{I(II)}=\omega^0_{sp}/\nu_\text{I(II)}$ with $\nu_\text{I(II)}$=1 or 2, and $\hbar\omega^0_{sp}=\epsilon_{sp}$, so the $\nu_\text{I(II)}$-photon resonant coupling was generated during the first (second) pulse duration.  $\tau_\text{I(II)}$ was set to the shaking period of $2\pi/\omega_\text{I(II)}$, satisfying the Floquet picture as well as ensuring the continuous changing of $\delta\omega_L$. 

In the rotating frame with frequency $\omega^0_{sp}=\nu_\text{I(II)}\omega_\text{I(II)}$, as shown in Eqs.~(S14) and (S18), the effective Bloch Hamiltonian of the system is expressed as $H_q(t)=-2\bar{t}_r\cos(q)\mathbb{I}+\bm{B}(q,t)\cdot \bm{\sigma}$ with
\begin{align}
\bm{B}(q,t)&=B_\rho(q,t)\hat{\bm{\rho}}(\theta(t)) + B_z(q)\hat{\bm{z}}, \nonumber 
\end{align}
where $\hat{\bm{\rho}}(\theta)=\cos(\theta)\hat{\bm{x}}+\sin(\theta)\hat{\bm{y}}$ and $B_z(q)=-2t_r \cos (q)$. The magnitude $B_\rho$($>$0) and the azimuthal angle $\theta$ of the transverse component of $\bm{B}(q,t)$ are determined by the lattice shaking parameters, $\{A_\text{I(II)},\nu_\text{I(II)},\varphi_\text{I(II)}\}$, in each pulse duration. Then, the system's evolution over a pulse duration $\tau$ is described by $R_{\hat{\bm{b}}}(\xi)=\exp(-i \xi \hat{\bm{b}}\cdot \bm{\sigma})$ with $\hat{\bm{b}}=\bm{B}/|\bm{B}|$ and $\xi=|\bm{B}|\tau/\hbar$. This operation corresponds to a rotation by $\xi$ around the $\hat{\bm{b}}$ axis on the Bloch sphere formed by the $|q,p\rangle$ and $|q,s\rangle$ states. Here the effect of the spin-independent term, $-2\bar{t}_r \cos(q)$, in $H_q$ is nullified by a gauge transformation. 

For the initial state of $(0,1)^\text{T}$, i.e., $|q,s\rangle$, the atom's final state after the Ramsey interferometry sequence is given by   
\begin{align}
\begin{pmatrix} c_p(q) \\ c_s(q) \end{pmatrix}&=
R_{\hat{\bm{b}}_\text{II}}(\xi_\text{II})R_{\hat{\bm{z}}}(B_z T_e/\hbar)R_{\hat{\bm{b}}_\text{I}}(\xi_\text{I})
\begin{pmatrix} 0 \\ 1 \end{pmatrix},
\end{align}
where $\hat{\bm{b}}_\text{I(II)}$ and $\xi_\text{I(II)}$ are the direction of $\bm{B}$ and the rotation angle for the first (second) lattice-shaking pulse, respectively. When $|B_z|\ll B_\rho$ and $\xi_\text{I}=\xi_\text{II} =\pi/4$, $\hat{\bm{b}}_\text{I(II)}\approx \hat{\bm{\rho}}_\text{I(II)}$, and the evolution matrix can be approximated as
\begin{align}
R_{\hat{\bm{\rho}}_\text{II}}&(\pi/4)R_{\hat{\bm{z}}}(B_z T_e/\hbar)R_{\hat{\bm{\rho}}_\text{I}}(\pi/4) \nonumber \\
	\approx&~\frac{1}{2} \Big[ e^{-i\frac{B_z T_e}{\hbar} \sigma_z}-e^{i(\frac{B_z T_e}{\hbar}+\theta_\text{I}-\theta_\text{II}) \sigma_z} \nonumber \\
	  &~~~-i\sigma_x\big(e^{-i(\frac{B_z T_e}{\hbar} +\theta_\text{I})\sigma_z}+e^{i(\frac{B_z T_e}{\hbar} -\theta_\text{II})\sigma_z}\big)\Big]. \nonumber
\end{align}
Then, the interferometric signal $n_p(q)=|c_p(q)|^2$ is given by 
\begin{equation}
n_p(q) = \frac{1}{2} \{ 1+\cos[2B_z(q) T_e/\hbar+\theta_\text{I}-\theta_\text{II}]\}.
\end{equation}

\begin{figure}
\includegraphics[width=8.0cm]{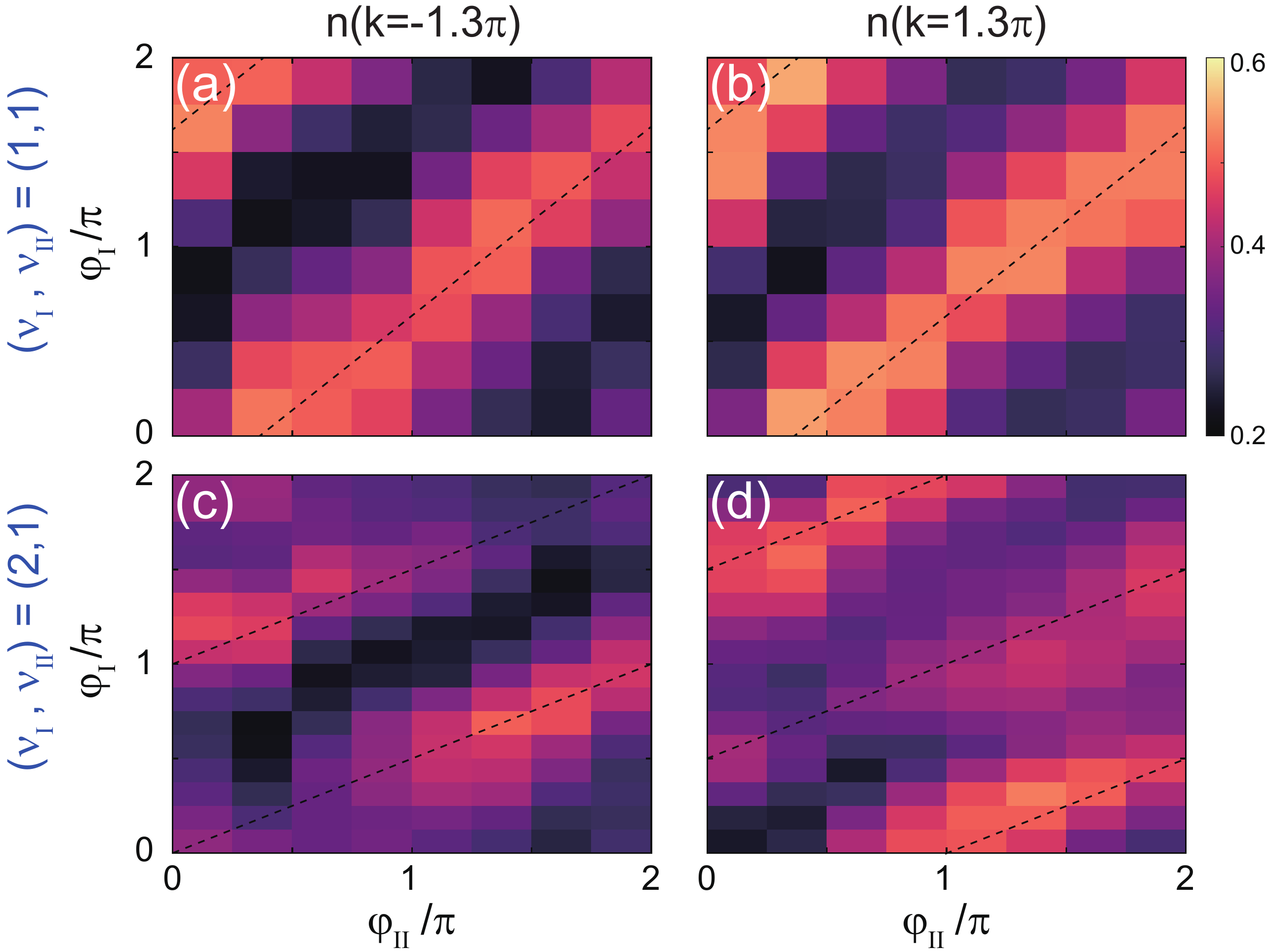}
\caption{\textbf{Driving phase dependence of the Ramsey fringe signal.} Atomic densities $n(k)$ measured at (a,c) $k=-1.3\pi$ and (b,d) $k=1.3\pi$ as functions of the driving phases $\varphi_\text{I}$ and $\varphi_\text{II}$ for $T_e=100~\mu$s. $\{\nu_\text{I},\nu_\text{II}\}=\{ 1,1\}$ in (a) and (b), and $\{2,1\}$ in (c) and (d). The experimental conditions are the same as those in Fig.~3. The dashed lines are guides to the eyes, indicating the maxima of the Ramsey fringes. In the case of $\{ \nu_\text{I},\nu_\text{II}\}=\{ 2,1\}$, the fring signal shows $\pi$-periodicity with increasing $\varphi_\text{I}$ as expected in Eq.~(S22).}
\end{figure}

In the case of $\{\nu_\text{I},\nu_\text{II}\}=\{1,1\}$ with $A_\text{I}=A_\text{II}=A$ and $\tau_\text{I}=\tau_\text{II}=\tau$, we have $B_\rho(q)=t_v$, $\theta_\text{I}=\varphi_\text{I}$, and $\theta_\text{II}=\varphi_\text{II}-\omega_{sp}^0(\tau+T_e)=\varphi_\text{II}-\omega_{sp}^0 T_e$ (mod $2\pi$)  [Eq.~(S14)]. Then Eq.~(S20) gives
\begin{equation}
n_p(q) = \frac{1}{2} \{ 1+\cos[\omega_{sp}(q) T_e+\varphi_\text{I}-\varphi_\text{II}]\}
\end{equation}
with $\hbar\omega_{sp}(q)=\epsilon_{sp}-4t_r \cos (q)$. 

In the case of $\{\nu_\text{I},\nu_\text{II}\}=\{2,1\}$,  $B_\rho(q)=2t_d |\sin (q)| $  and $\theta_\text{I}=2\varphi_\text{I}+\text{sgn} (q) (\pi/2)$ for the first pulse [Eq.~(S18)], and $B_\rho(q)=t_v$ and $\theta_\text{II}=\varphi_\text{II}-\omega_{sp}^0 T_e$  for the second pulse. Although the assumption of $|B_z|\ll B_\rho$ is not valid near $q=0$ and $\pi$, we see that Eq.~(S20) results in
\begin{equation}
n_p(q) = \frac{1}{2} \{ 1-\text{sgn}(q)\sin[\omega_{sp}(q)T_e+2\varphi_\text{I}-\varphi_\text{II}]\}.
\end{equation}
Because of the $\text{sgn(q)}$ factor, the Ramsey fringe signal is asymmetric for opposite momenta $\pm q$. Another noticeable feature is that the dependence of $n_p(q)$ on $\varphi_\text{I}$ is different from that in the $\nu_\text{I}=1$ case, so the interference signal has $\pi$ periodicity rather than $2\pi$ with increasing $\varphi_\text{I}$~[Fig.~3(f) inset and Fig.~S1].

\begin{figure}
\includegraphics[width=8.4cm]{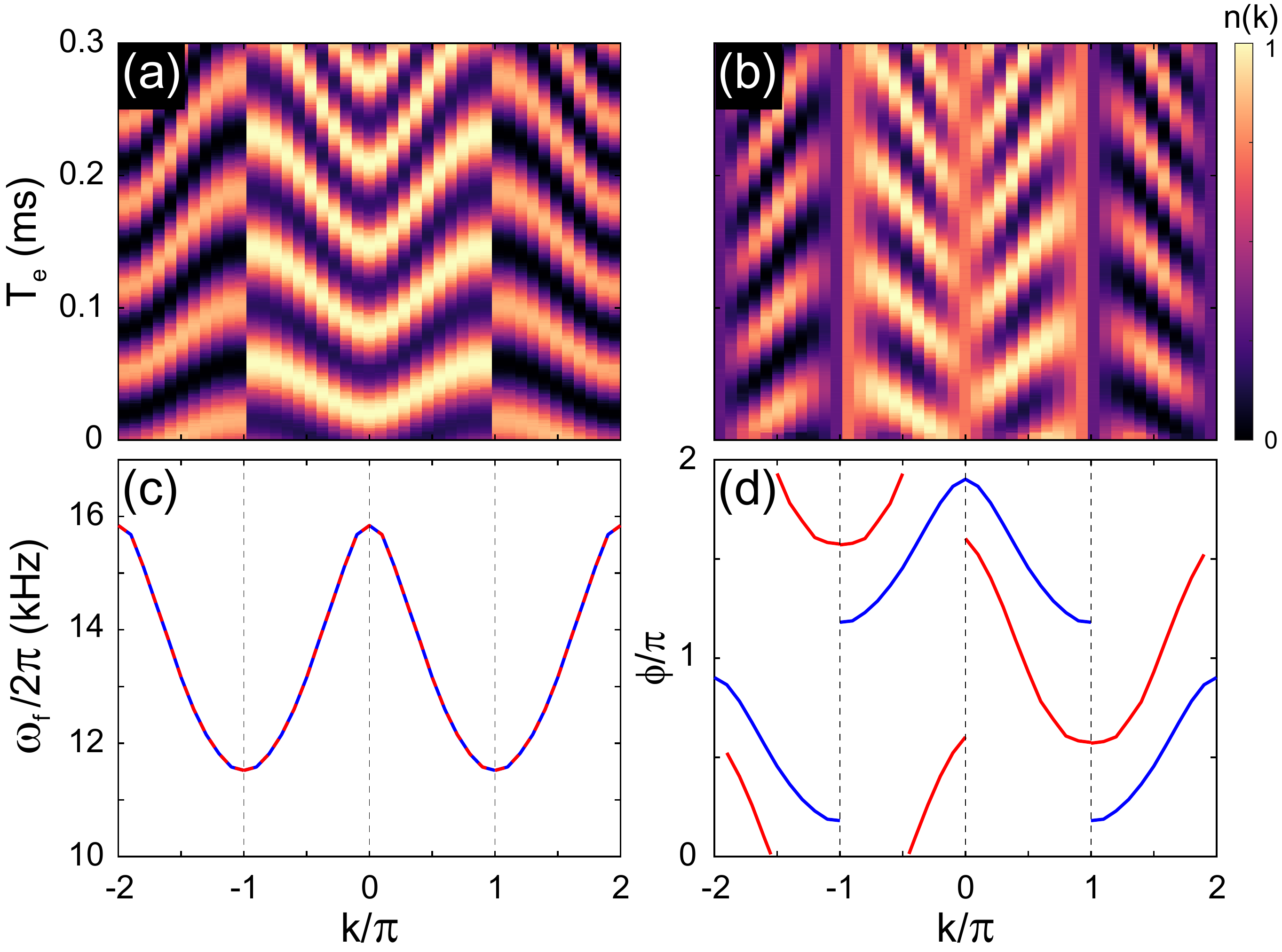}
\caption{\textbf{Numerical simulation of the Ramsey interferometry.} Momentum-resolved Ramsey interferometry signals for (a) $\{\nu_\text{I},\nu_\text{II}\} =\{1,1\}$ and (b) $\{2,1\}$, calculated from Eq.~(S19). The lattice and shaking parameters are set to the experimental condition in Fig.~3. (c) Oscillation frequency $\omega_f$ and (d) phase $\phi$ of the interferomety signal as functions of momentum $k$. The blue and red lines denote the values for $\{\nu_\text{I},\nu_\text{II}\}=\{1,1\}$  and $\{2,1\}$, respectively.}
\end{figure}

\begin{figure}
\includegraphics[width=8.4cm]{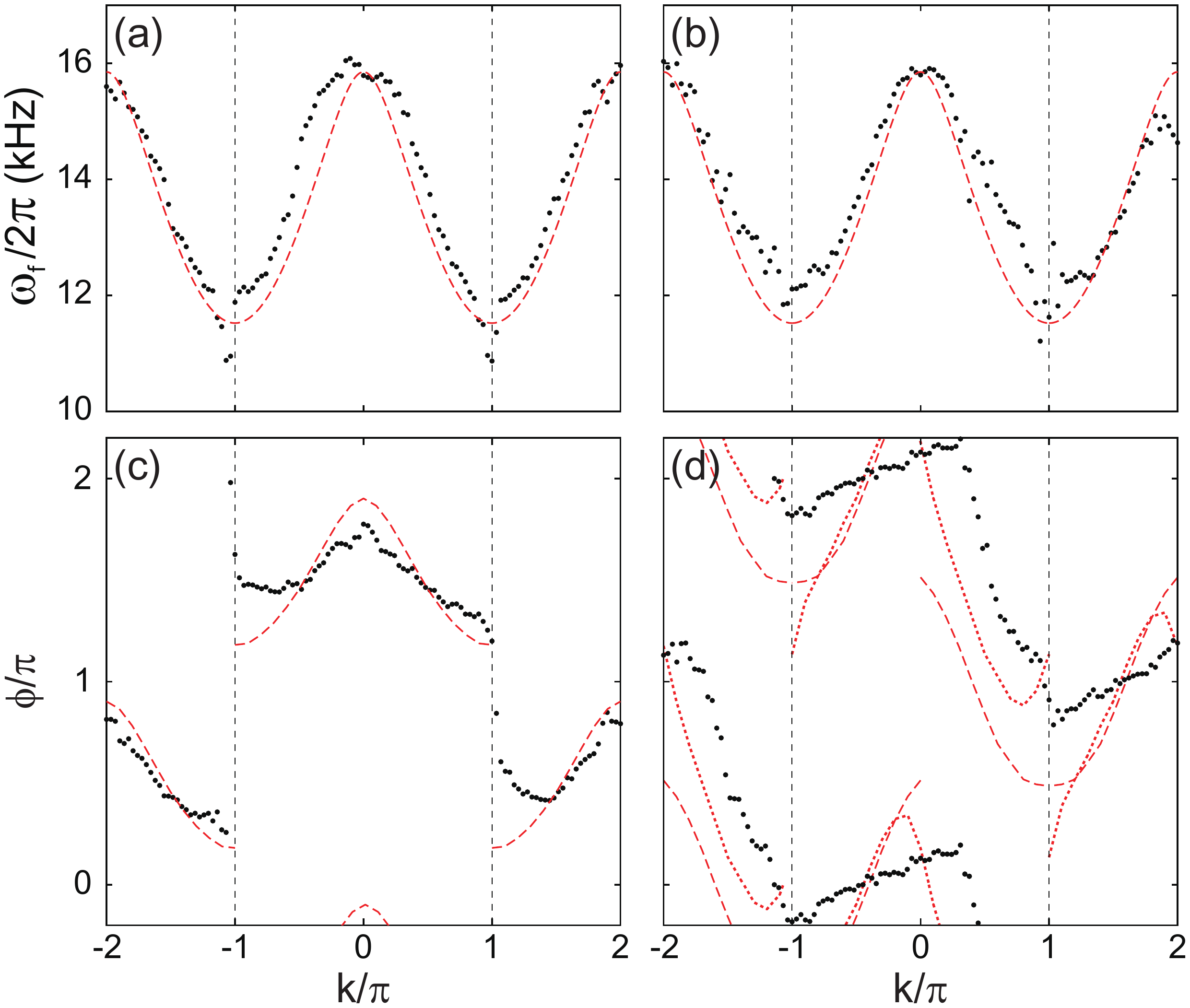}
\caption{\textbf{Characterization of the measured Ramsey fringe signals.} From a sinusoidal function fit to the experimental data in Fig.~3(c) and 3(d), the oscillation frequency $\omega_f(k)$ and phase $\phi(k)$ are determined, and they are displayed, respectively, in (a) and (c) for $\{\nu_\text{I},\nu_\text{II}\} =\{1,1\}$ [Fig.~3(c)], and (b) and (d) for $\{\nu_\text{I},\nu_\text{II}\} =\{2,1\}$ [Fig.~3(d)]. The red dashed lines indicate the numerical results in Fig.~S2. In (d), the red dotted lines shows the numerical result for the case where an additional uniform transverse field is included for the first pulse period (see text).}
\end{figure}


\subsection{Numerical simulation of the Ramsey interferometry}

Using Eq.~(S19), we numerically calculated the Ramsey interferometry signals for our experimental conditions. The results for $\{\nu_\text{I},\nu_\text{II}\}=\{1,1\}$ and $\{2,1\}$ are shown in Fig.~S2(a) and S2(b), respectively. The oscillations at each $k$ is characterized by a sinusoidal curve, $n(k,T_e)\propto \sin[\omega_f(k)T_e+\phi(k)]$ with $\omega_f(k) = \omega_{sp}(k)$ [Fig.~S2(c)]. We observe $\phi(k)-\phi(-k) =0$ for $\{\nu_\text{I},\nu_\text{II}\}=\{1,1\}$ and $|\phi(k)-\phi(-k)|=\pi$ for $\{\nu_\text{I},\nu_\text{II}\}=\{2,1\}$  [Fig.~S2(d)], which are consistent with the expectations from Eqs.~(S21) and (S22).

In Fig.~S3, we compare the experimental data in Fig.~3(c) and 3(d) with the numerical results. Here the measured spectra are characterized by fitting a sinusoidal curve with frequency $\omega_f(k)$ and phase $\phi(k)$ to the oscillating signal at each $k$. The obtained $\omega_f(k)$ is found to be in good agreement with the  calculated $\omega_{sp}(k)$ [Fig.~S3(a) and S3(b)]. In  the $\{\nu_\text{I},\nu_\text{II}\} =\{1,1\}$ case, the oscillation phase $\phi(k)$ is also found to be well described by the numerical results~[Fig.~S3(c)]. The small deviations near $k=\pm\pi$ might be attributed to the imperfection of the band mapping technique or some systematic effects in our experiment. 

On the other hand, in the $\{\nu_\text{I},\nu_\text{II}\} =\{2,1\}$ case, we observe that the measured $\phi(k)$ shows a non-negligible deviation from the numerical result [Fig. S3(d)], although it reasonbly demonstrates the asymmetric property of $|\phi(k)-\phi(-k)|=\pi$. The deviation is most pronounced near $k=0$, where it should be technically difficult to determine $\phi$ because of small oscillation amplitude. In order to understand the measurement result, it might be necessary to take into account the one-photon off-resonant effect that is neglected in our description. As an attempt, we examined a situation where an additional transverse field $B_x$ is added along the $\hat{\bm{x}}$ direction for the first pulse period to mimic the off-resonant effect. We observed that with $B_x\sim B_\rho/3$, the numerically recalculated fringe signal exhibits similar deviating behavior to the experimental result [Fig.~S3(d)].

\subsection{Two-tone driving scheme}

In this section, we describe the two-tone driving scheme where the optical lattice is driven as 
\begin{align}
\delta\omega(t) = A \sin(\omega t)+\tilde{A}\sin(2\omega t+\varphi).
\end{align}
Following the derivation in Sec.~A, we obtain the Hamiltonian in the comoving frame as 
\begin{align}
H =& \frac{p^2}{2m}+\frac{V_L}{2}\cos(2k_L x)-d \omega \sin(\omega t)p \nonumber \\
     &-2\tilde{d}\omega \sin(2\omega t+\varphi)p.
\end{align}
In two-band tight binding approximation, the Hamiltonian $H'$ in a rotating reference frame with frequency $2\omega$ is given by
\begin{equation}
H' = \sum_j \Big\{\Psi_j^\dagger K'(t) \Psi_j -(\Psi_j^\dagger J'(t) \Psi_{j+1}+\text{H.c.})\Big\}\nonumber,
\end{equation}
with
\begin{align}
K'(t) =& \Delta_2 \sigma_z +h_0^{sp}\sin(\omega t)R(-2\omega t)\sigma_y \nonumber \\
		&+\tilde{h}_0^{sp}\sin(2\omega t+\varphi)R(-2\omega t)\sigma_y, \nonumber\\
J'(t) =&~~\bar{t}_r \mathbb{I}+ t_r \sigma_z \nonumber \\
	&-i[\bar{h}_1 \sin(\omega t)+\bar{\tilde{h}}_1 \sin(2\omega t+\varphi)] \mathbb{I} \nonumber \\
	&-i[h_1 \sin(\omega t)+\tilde{h}_1 \sin(2\omega t+\varphi)] \sigma_z \nonumber \\
	&-h_1^{sp}\sin(\omega t)R(-2\omega t) \sigma_y \nonumber \\
	&-\tilde{h}_1^{sp}\sin(2\omega t+\varphi)R(-2\omega t) \sigma_y.
\end{align}
Here the tilde symbol indicates that the elements are derived from the $2\omega$ driving with the amplitude $\tilde{A}$. 

\begin{figure}[t]
\includegraphics[width=8.4cm]{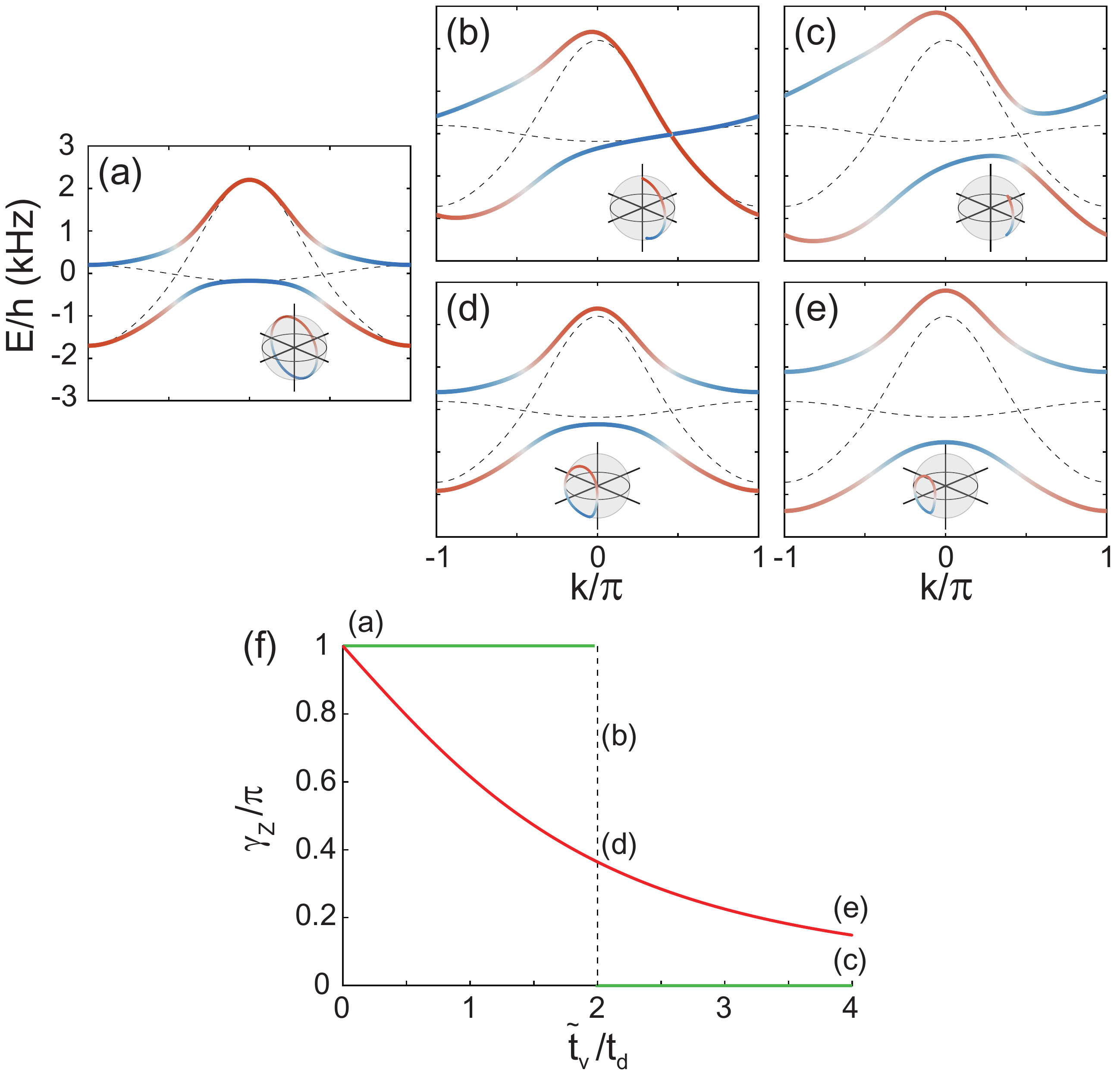}
\caption{\textbf{Topological phase transitions of the Creutz ladder system.} Floquet band structures calculated from Eq.~(S27) for various driving conditions: $2\omega = \epsilon_{sp}/\hbar$, $A/2\pi = 8$~kHz, and (a) $\{\tilde{t}_v/t_d, \varphi\}=\{0,\pi/2 \}$, (b) $\{2,\pi/2\}$, (c) $\{4,\pi/2\}$, (d) $\{2,0\}$, and (e) $\{4,0\}$. The insets show the Bloch vector trajectory of the ground band. (f) Zak phase $\gamma_Z$ of the ground band as a function of $\tilde{t}_v/t_d$ for $\varphi=\pi/2$ (green line) and $\varphi=0$ (red line). For $\varphi=\pi/2$, a topolgical phase transition occurs at $\tilde{t}_v/t_d = 2$ with a sudden change of $\gamma_Z$ and the band gap closing shown in (b).}
\end{figure}

The effective Hamiltonian, $H_\text{eff}$, of this peridocially driven system can be obtained in a same manner shown in Sec.~B using the high frequency expansion method. Here, it is important to note that the Fourier components, $H_{\pm 1}$ and $H_{\pm 3}$ are generated only by the $\omega$ driving to have the same forms expressed in Eq.~(S15), and $H_{\pm 2}$ and $H_{\pm4}$ are generated only by the $2\omega$ driving, resulting in the same forms of $H_{\pm1}$ and $H_{\pm2}$ in Eq.~(S11). Then, from Eq.~(S9), we can see that the effects of the two drivings arise additively in the system, thus leading to 
\begin{align}
H_{\text{eff}} \approx &  \sum_j  \Psi_j^\dagger \bigg[\Delta_2\sigma_z+\tilde{t}_v R(\varphi)\sigma_x\bigg]\Psi_j     \nonumber \\
& -\sum_j \bigg[ \Psi_j^\dagger(\bar{t}_r\mathbb{I} +t_r\sigma_z-t_d i\sigma_y)\Psi_{j+1}+\text{H.c.}\bigg]
\end{align}
with $\tilde{t}_v = \tilde{h}_0^{sp}/2$ and $t_d =h_0^{sp}h_1/2\hbar\omega$. Here the same approximations used in Eqs.~(S13) and (S17) are applied. Accordingly, in the quasimomentum space, the Bloch Hamiltonian is given by 
\begin{equation}
H_q = -2\bar{t}_r\cos(q)\mathbb{I}+\bm{B}(q) \cdot \bm{\sigma}\nonumber
\end{equation}
with
\begin{align}
\bm{B}(q) &= \tilde{t}_v \hat{\bm{\rho}}+ 2 t_d\sin(q) \hat{\bm{y}}+ [\Delta_2-2t_r\cos(q)]\hat{\bm{z}} \nonumber \\
\hat{\bm{\rho}}&=\cos(\varphi)\hat{\bm{x}}+\sin(\varphi)\hat{\bm{y}}. 
\end{align}

In this two-tone driving scheme, the topological phase transitions of the Creutz ladder system can be studied. 
For example, when $\varphi = \pi/2$ and 
\begin{align}
\bm{B}(q) =&~ [\tilde{t}_v + 2t_d\sin(q)] \hat{\bm{y}}+ [\Delta_2-2t_r\cos(q)]\hat{\bm{z}} \nonumber, 
\end{align}
the system undergoes a topological phase transition as $\tilde{t}_v/t_d$ changes across the critical point of $\tilde{t}_v/t_d=2$. At the critical point, the band gap closes at $q_c$ such that $\Delta_2-2t_r\cos(q_c)=0$ giving $|\bm{B}(q_c)|=0$ [Fig.~S4(b)], and  the Zak phase $\gamma_Z$ suddenly changes by $\pi$ [Fig.~S4(f)]. The Zak phase is numerically calculated as $\gamma_Z = i \int_{BZ} dq \langle u_q |\partial_q|u_q\rangle$, where $u_q$ is the Bloch function of the band~\cite{Zak89}. 

On the other hand, when $\varphi=0$, i.e., the two transverse fields from the $\omega$ and $2\omega$ drivigns are orthogonal to each other, 
\begin{align}
\bm{B}(q) =&~ \tilde{t}_v \hat{\bm{x}} + 2t_d\sin(q) \hat{\bm{y}}+ [\Delta_2-2t_r\cos(q)]\hat{\bm{z}}.
\end{align}
In this case, $\gamma_Z$ can change continuously with varying $\tilde{t}_v/t_d$ [Fig.~S4(f)], and the system can make a topological phase transition without having band gap closing [Figs.~S4(a,d,e)].


\begin{references}
\bibitem{footnote1} The long-range tunneling in the $p$ band is included in our calculation of $\omega_{sp}(q)$.
\end{references}
\end{document}